\documentclass[aps,preprint,showpacs,superscriptaddress,groupedaddress]{revtex4}  
\usepackage{graphicx}  
\usepackage{dcolumn}   
\usepackage{bm}        
\usepackage{amssymb}   
\usepackage{amsmath}
\begin{document}

\title{Effect of liquid state organization on microstructure and strength of model multicomponent solids}

\author{ Kulveer Singh }
\email{kulveersingh85@gmail.com}
\author{Yitzhak Rabin}%
 \email{yitzhak.rabin@biu.ac.il}
\affiliation{Department of Physics, and Institute of Nanotechnology and Advanced Materials, \\Bar-Ilan University,
Ramat Gan 52900, Israel}%

\date{\today}

\begin{abstract}
	When a multicomponent liquid composed of particles with random interactions is slowly cooled below the freezing temperature, the fluid reorganises in order to increase (decrease) the number of strong (weak) attractive interactions and  solidifies into a microphase-separated structure composed of domains of strongly and of weakly interacting particles. Using Langevin dynamics simulations of a model system we find that the limiting tensile strength of such solids can exceed that of one-component solids.  
\end{abstract}

\maketitle



Saturation in mechanical performance enhancement of single-component-based materials shifted the attention of  material scientists and engineers to multicomponent alloys for improving mechanical properties of solids \cite{Gludovatz2014, Ye2016, Yang2018}. Understanding the relation between limiting strength (yield stress, fracture) and the microscopic details of multicomponent structures is of great interest due to their technological implications in developing new materials with desirable properties. 
Following theoretical investigations of the thermodynamics of multicomponent liquids \cite{Stell1982, Cates1998, Evans1999, Fasolo2005, Frenkel2013, Bagchi2013}, recently there have been studies of both static and dynamic properties of model systems in which all particle sizes are identical but all interactions between particle pairs are different (AID) \cite{Lenin2015, Lenin2016, Tanaka2016, Itay2019}. 
It was demonstrated that, as the system is cooled towards the freezing temperature, microphase separation occurs in AID fluids which is characterized by neighborhood identity ordering (NIO) where particles tend to favor neighbors with whom they share strong attractive interactions and avoid neighbors with whom they have weaker interactions. NIO progressively increases as temperature is further reduced until the system freezes. It was found that even though an AID liquid freezes into a crystalline state which is similar to that of a one-component solid in terms of spatial ordering of the particles,  it does not reach equilibrium with respect to NIO and forms an AID glass \cite{Dino2016}.  Note that AID glasses are quite different from amorphous glasses \cite{Nagel1996, Frank2001, Peter1999, Pinaki2005, Zippelius2005, Raza2015}  in which rapid cooling (or size polydispersity) results in non-crystalline (amorphous) arrangement of the centres of mass.  

How do mechanical properties of such AID glasses differ from those of a one-component (1C) solid of the same crystal structure and average nominal interaction strength between particles (i.e., mean of the AID interaction parameter distribution)? One expects the linear elastic response to small applied strain to be quite similar for the two materials since it depends only on their average properties. However, because of the presence of weak bonds in AID (compared to 1C) solids, one may expect that these bonds will break at a lower critical stress than bonds of average strength and, therefore, that the limiting strength of AID solids is lower than that of 1C solids. In this work we show that limiting strength and mode of fracture of AID solids depends on the history of their preparation prior to complete freezing and that increasing the degree of NIO may enhance the elastic moduli and the limiting strength of AID solids beyond that of their 1C counterparts.

We carry out Langevin dynamics simulations using LAMMPS  \cite{Plimpton1995}, to investigate the deformation and fracture of 2D solid materials prepared by cooling a dense AID liquid. Our simulation box contains $N = 2400$ particles of mass $m$ and diameter $\sigma$ with density $\rho = 0.9$ ($\rho$=N/area). Periodic boundary conditions used during system preparation, prior to uniaxial loading. Particles interact with each other via a short-range Lennard-Jones (LJ) potential given by $U_{ij}(r)=4\epsilon_{ij}[(\sigma/r)^{12}-(\sigma/r)^{6}]$ which is cut and shifted to zero at cut-off distance of $2.5\sigma$. The interaction strength $\epsilon_{ij}$ is randomly chosen from a uniform distribution in the range $1-4$ for AID system and $\epsilon_{ij}=2.5\epsilon$ (mean of AID distribution) is assigned to 1C system.  The motion of the particles in the system is described by the  Langevin equation 
\begin{eqnarray}
m\ddot{\bf{r}}_i(t) = -\frac{\partial U}{\partial {\bf{r}}_i} -\zeta \dot{\bf{r}}_i(t) + \eta_i(t)
\label{eq:langevin},
\end{eqnarray}
which contains the total potential energy $U$ (sum over all $U_{ij}$), particle friction coefficient $\zeta$ and random thermal force $\eta_i$ with magnitude proportional to $(\zeta k_BT/\Delta t)^{1/2}$, where $k_B$, $T$ and $\Delta t$ are Boltzmann's constant, temperature and integration time step, respectively.  All physical quantities are expressed in reduced LJ units with mass $m$, particle diameter $\sigma$, interaction parameter $\epsilon$ and $k_B$ set to $1$. This gives the LJ time unit, $\tau_{LJ} = (m\sigma^2/\epsilon)^{1/2}=1$ and the integration time-step $\Delta t$ is taken as $0.005$. The systems are prepared with $\zeta = 0.02$ which corresponds to damping time $\tau_d=50$, but during uniaxial loading simulations we used a larger value of the friction, $\zeta = 50$  ($\tau_d=0.02$), in order to ensure fast relaxation of the temperature and suppress stress-induced  heating of the system. NIO in the AID system is characterized by  $\langle \epsilon_i^{eff} \rangle$ where \cite{Lenin2015}
\begin{eqnarray}
 \epsilon_i^{eff} = \sum_{j=1}^{n_b}\epsilon_{ij}/n_b
\label{eq:nio},
\end{eqnarray}
is the effective interaction parameter of particle $i$ having $n_b$ neighbours. 

\begin{figure}[ht] 
\includegraphics[width=\linewidth]{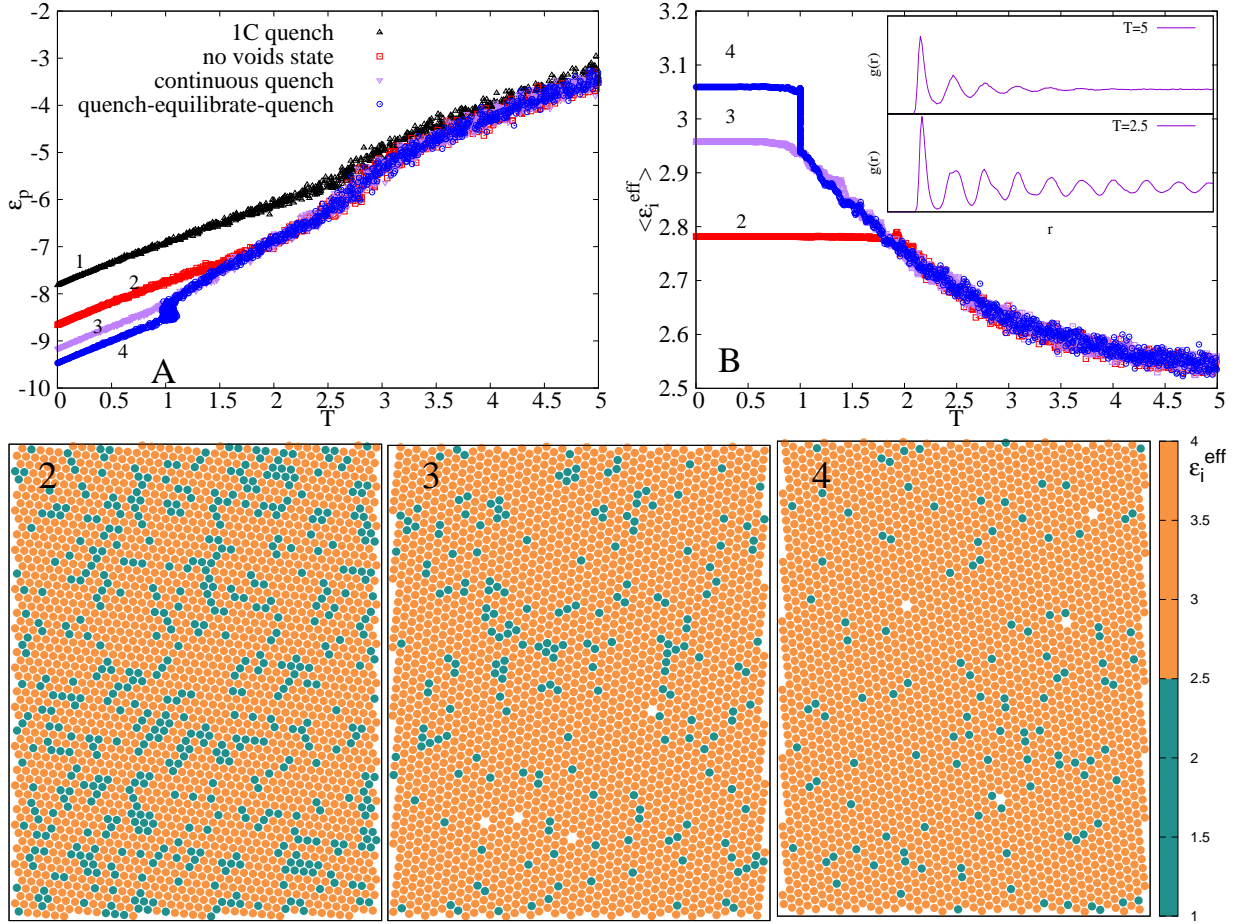}
	\caption{\label{fig:cooling} (A) shows potential energy per particle $\epsilon_p$ vs temperature of 1C (curve 1) and AID (curves 2, 3, 4) systems. When the AID system is slowly cooled to $T=0$ with $dT/dt = 10^{-5}/\tau_{LJ}$, it freezes into a crystal with no defects (curve 2) or with defects (curve 3). Curve 4 represents an AID system slowly cooled to $T=1$, equilibrated at $T=1$ and then cooled to $T=0$. (B)  shows $\langle \epsilon_i^{eff}\rangle$ vs T for curves 2, 3, and 4. Inset in (B) shows the radial distribution function of AID system at $T= 5$ and $2.5$. Typical snapshots corresponding to curves 2, 3 and 4 are shown in the lower panels. Particles are colored according to $\epsilon_i^{eff}$ values.}
\end{figure}

The AID system was prepared by randomly generating a set of $N(N-1)/2$ values of $\epsilon_{ij}$ in the range $1-4$, putting the particles on a square lattice,  equilibrating the AID liquid at high temperature ($T=5$) and then quenching the system to $T=0$ with $dT/dt = 10^{-5}/\tau_{LJ}$. Figure \ref{fig:cooling}A shows the cooling curves of the AID system where we have plotted the average potential energy per particle ($\epsilon_p$) as a function of temperature and compared it with the 1C system. The liquid to solid transition (liquid-solid coexistence) takes place over a broad temperature range around $T\sim 2.5$, as indicated by appearance of long-range order in the radial distribution function $g(r)$ (see inset of Fig.\ref{fig:cooling}B) \cite{Gasser2009}. Below this temperature, $\epsilon_p$ is lower in AID than in 1C system (compare curve 1 with curves 2, 3 and 4 in Fig. \ref{fig:cooling}A), a consequence of NIO in the former but not in the latter system. To confirm this conclusion we computed $\langle \epsilon_i^{eff} \rangle$ as a function of T (Fig. \ref{fig:cooling}B) and indeed NIO increases and eventually saturates as T decreases (snapshots of the system at different temperatures are shown in Fig. S1\cite{SI}).   Depending on initial conditions, the resulting AID solid can be either defect-free (snapshot 2 in Figure \ref{fig:cooling}, lower panel) or contain defects (snapshot 3 in Figure \ref{fig:cooling}, lower panel). Only the latter situation is observed in 1C solids, independent of initial conditions. Importantly, below $T\sim 2$, the presence of defects lowers the potential energy and increases the NIO compared to the defect-free system (see  Figs. \ref{fig:cooling}A and B). This phenomenon can be attributed to the mobility of defects (voids) in the range of temperatures at which solid and liquid phases coexist, which provides a mechanism for particle rearrangement that tends to increase NIO and reduce the energy of the system (see movie M1 \cite{SI}). This continues until the system completely freezes and defects are immobilized at temperatures below $T\sim 1$, resulting in saturation of $\langle \epsilon_i^{eff} \rangle$ (Fig. \ref{fig:cooling}B). In order to further improve the NIO of the AID solid we prepared the system using the quench-equilibrate-quench method. First, we slowly quenched the system to $T=1$, equilibrated it  at this temperature for $10^5 \tau_{LJ}$ and then slowly quenched it again from $T=1$ to $T=0$ (see curve 4 and snapshot 4 in Fig. \ref{fig:cooling}). In this way, we were able to reach $\langle \epsilon_i^{eff} \rangle >3.05$, substantially higher than the mean value of $2.5$ of the uniform distribution that is approached at high temperatures. Further improvement of NIO by equilibration at yet lower temperature is limited by the fact that defect mobility decreases dramatically below this temperatures and much longer simulations would be necessary. 

\begin{figure}[h]
\includegraphics[width=\linewidth]{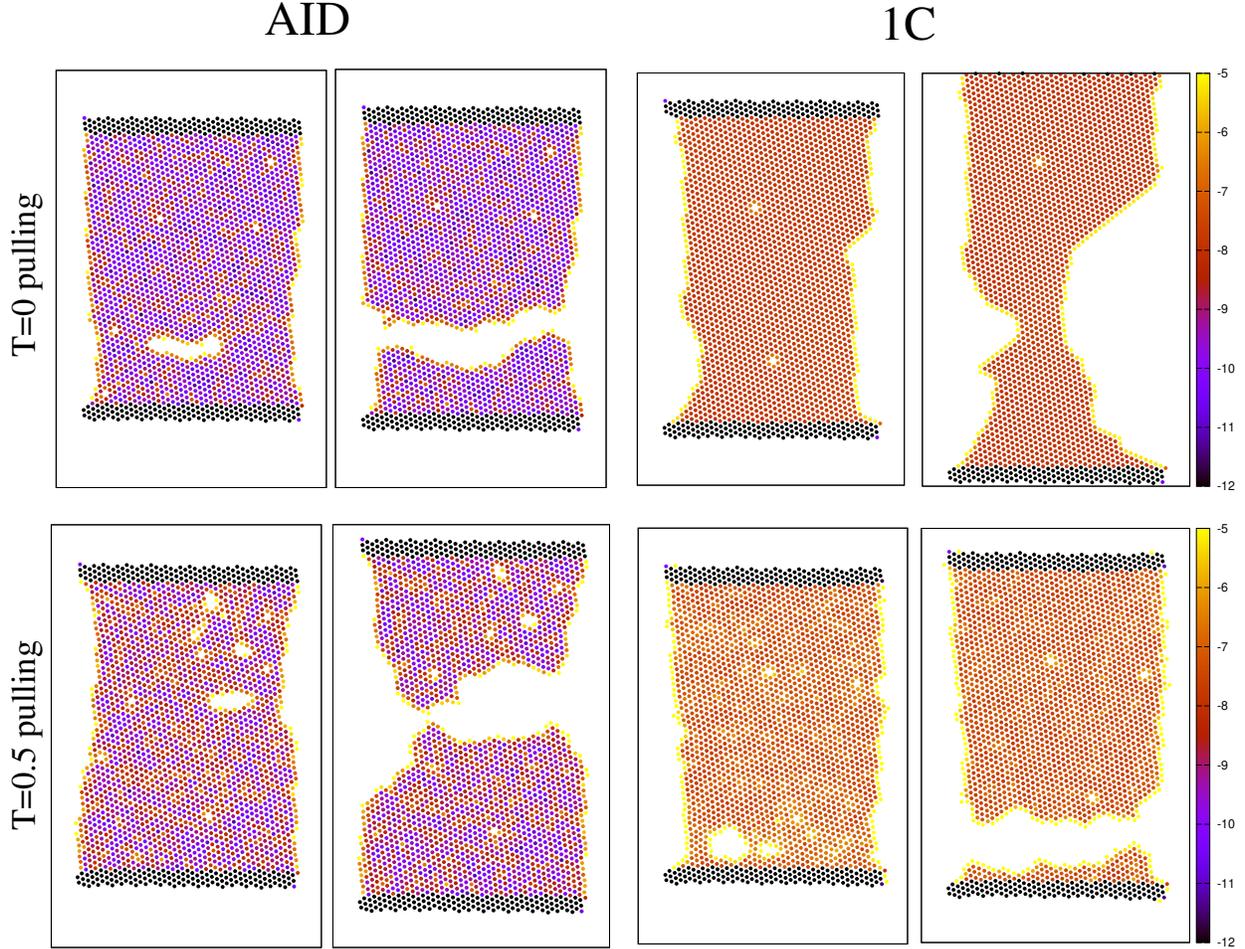}
	\caption{ \label{fig:snapshots} Typical snapshots of fracture mode of AID and 1C system during uniaxial loading simulation done at $T=0$ (upper panels) and $T=0.5$ (lower panels). Initial spatial configuration for both systems corresponds to preparation by quench-equilibrate-quench method (case 4 in Fig \ref{fig:cooling}). Particles are colored according to potential energy. }
\end{figure}

Uniaxial loading simulations were performed by removing the periodic boundary conditions and  equilibrating the solid at the loading temperature. After equilibration, four boundary layers of particles on the top and bottom of the solid (black particles in Figure \ref{fig:snapshots}) were frozen and pulled with constant velocity $v= 0.0001$ (in units of $\sigma/\tau$) in opposite directions along the y axis. The positions of the particles in the top and bottom layer are modified using  $y(t) = y(0) + vt$ , where $y(0)$ is the initial y coordinate of the particle. Strain is measured by computing the distance between the centers of mass of particles in the lower row of the top layer and upper row of the bottom layer. The instantaneous virial stress is computed by taking the $yy$ component of the stress tensor\cite{Yip1991} 
\begin{eqnarray}
 {\bf\Gamma}=\Big(\sum_i m {\bf v_i}\otimes {\bf v_i} + \frac{1}{2} \sum_{j\neq i} {\bf r_{ij}}\otimes {\bf F_{ij}}\Big)/V
\label{eq:stress}.
\end{eqnarray}
Here ${\bf v_i}$ is velocity of particle $i$, $\bf{F}_{ij}$ is the force acting on particle $i$ due to particle $j$, $\bf{r}_{ij}=\bf{r}_i - \bf{r}_j$, where $\bf{r}_i$ and $\bf{r}_j$ are the positions of $i$ and $j$ particles, and $V$ is the volume of the system.

\begin{figure*}
\includegraphics[width=\linewidth]{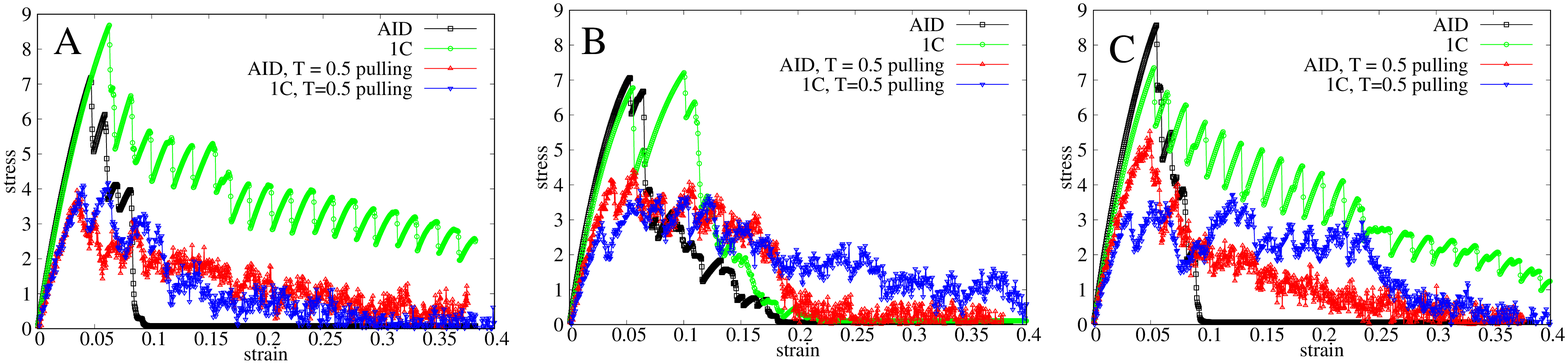}
	\caption{\label{fig:slow_quench} (A), (B) and (C) represents stress-strain curve for three different initial configurations corresponding to continuous cooling without defects, continuous cooling with defects and quench-equilibrate-quench methods of preparation represented by curves 2, 3 and 4 in Fig. \ref{fig:cooling}, respectively. Uniaxial loading simulation is performed at $T=0$ and $T=0.5$. 1C system used for comparison has exactly same initial spatial configuration as quenched AID system. }
\end{figure*}

Figure \ref{fig:snapshots} shows two snapshots of AID and 1C solids taken at different times during uniaxial loading at $T=0$ (upper panels) and $T=0.5$ (lower panels).  The initial configuration of both systems is shown in snapshot 4 of Fig. \ref{fig:cooling} (quench-equilibrate-quench preparation) where, in the case of 1C we replaced all $\epsilon_{ij}$  by $2.5$ in order to have same initial number and locations of defects in both types of solids. When the uniaxial loading simulation was performed at $T=0$, we observed brittle and plastic fracture modes in AID and 1C solids, respectively (brittle fracture was also observed in simulations of ordinary two component glasses \cite{Langer1998, Barrat2004, Rodney2009, karmakar2011, Bouchbinder2018}).  When the AID solid is strained, new voids are nucleated and, as strain is further increased, these voids coalesce into cracks and eventually fracture results (see Fig. \ref{fig:snapshots} and movie M2\cite{SI}). When the 1C solid is strained, a neck begins to form (without nucleation of new voids) and becomes narrower by sliding of particles along the symmetry lines of the hexagonal crystal, as strain is further increased (see Fig. \ref{fig:snapshots} and movie M3\cite{SI}). 1C solids undergo large plastic deformations prior to fracture in most of the simulations we performed (for different initial configurations). We would like to mention here that the degree of plasticity in both AID and 1C solids depends on the number of defects, their location and also on the orientation of the crystal symmetry lines with respect to the pulling direction. When the loading simulation is performed at $T=0.5$, the deformation of both AID and 1C solids is accompanied by nucleation and coalescence of voids into cracks and eventually results in fracture (see Fig. \ref{fig:snapshots}). 

In  Figure \ref{fig:slow_quench}, we present stress versus strain plots for AID solids with different degrees of NIO (corresponding to curves 2, 3 and 4 of Fig. \ref{fig:cooling}B ), and for their 1C counterparts (with the same initial spatial configuration as the corresponding AID solid). We find that the Young modulus E of AID solids depends on the method of preparation and on the final temperature and monotonously increases with degree of NIO: $E_4:E_3:E_{1C} = 219: 211: 193$  and $130:122:112$ where the subscripts refer to the systems corresponding to snapshots $4$ (prepared by quench-equilibrate-quench) and $3$ (continuous cooling) in Figure \ref{fig:cooling}, for solids probed at $T=0$ and $T=0.5$, respectively (Fig. S2 \cite{SI}). In case of $T=0$ loading, as the degree of NIO increases, so does the critical stress of the AID solid and eventually it becomes larger than that of the corresponding 1C solid. This happens because as NIO increases, the number of strongly interacting neighbors ($\epsilon_{ij}>2.5$) increases and that of weakly interacting neighbors ($\epsilon_{ij}<2.5$) decreases. Both effects suppress coalescence of existing voids and formation of new ones and increase the ability of the solid to sustain higher stress. We observe that while the limiting tensile strength of 1C solids is only weakly-dependent on conditions of preparation, the degree of plasticity depends strongly on preparation through its effect on the distribution of defects (compare Figs. \ref{fig:slow_quench}B and C). Furthermore, while defects always decrease the tensile strength of 1C solids (compare Figure \ref{fig:slow_quench}A with Figure \ref{fig:slow_quench}B, C), the tensile strength of AID solids increases with the degree of NIO which has non-monotonic dependence on the concentration of defects (recall that the presence of a moderate number of mobile defects during preparation of the AID solid increases NIO). During $T=0.5$ loading, voids proliferate in both AID and 1C solids and fracture proceeds via coalescence of these voids (see movies M4 and M5\cite{SI}). The tensile strength of both AID and 1C solids is strongly reduced compared to $T=0$ loading, but its  decrease is smaller in AID than in 1C solids. Similarly to the case of $T=0$ loading, the tensile strength of AID solids increases with increasing NIO (compare Figs. \ref{fig:slow_quench}).  Another interesting observation is that solids (either AID or 1C) that undergo plastic deformation under $T=0$ loading, become more brittle under $T=0.5$ loading. Conversely, solids that are more brittle under $T=0$ loading, tend to undergo plastic deformation under $T=0.5$ loading. We would like to mention that we compared uniaxial loading of 1C solids prepared as described above, to that of 1C solids prepared by slow cooling of a 1C liquid, and did not find substantial differences in the response of two systems (not shown). Note that the above comparison between AID and 1C solids refers to 1C systems with nominal interaction parameter of $2.5$; if however, we compare an AID solid with $\langle\epsilon_i^{eff}\rangle=3.08$ to 1C solid with $\epsilon_{ij}=3.08$, the critical stress of the latter is going to be slightly higher that of the former (not shown). 

In addition to the above slow cooling results, we studied the uniaxial loading response of solids prepared by fast cooling ($dT/dt = 0.1$) of the AID liquid, and compared it with the corresponding 1C material (with the same crystal structure and defects at $T=0$). Fast cooling leads to the formation of many crystallites with different orientations that are separated by defect-enriched grain boundaries.  Both AID and 1C solids exhibit large reduction in limiting tensile strength as compared to solids formed by slow cooling (not shown). We performed loading simulations for different initial configurations of rapidly cooled systems and observed that the limiting tensile strength and fracture behavior do not depend on the characterstics of the constituent particles but strongly depend on the structure of the grain boundaries. When these materials are strained, a crack originates at the grain boundary, propagates along it upon further straining and finally leads to  fracture of the solid (see Fig. S3 \cite{SI}).   


In this work we used the AID model to study the effects of preparation from the liquid state on structural and mechanical properties of multicomponent solids. While this model is clearly an idealization (all particles are different and the interaction is assumed to be of Lennard-Jones type) and the present study is limited to 2D, we believe that it captures, albeit qualitatively, some of the important properties of high entropy alloys and solid solutions \cite{Cantor2004, Yeh2004,Zhang2014}, such as their ability to undergo self-organization upon cooling in the liquid state. We have quantified this self-organization on the nanoscale (NIO) in terms of local and average effective interaction parameters and proposed a cooling algoritm that optimizes the cohesive properties of the resulting solid. We demonstrated that NIO enhaces the limiting strength of AID solids under uniaxial loading, by increasing the critical stress for fracture and by suppressing plastic deformation compared to one-component crystalline solids. Finally, we found that the tensile strength of AID solids is less affected by increasing temperature than that of their one-component counterparts. We believe that our model provides important insights about the relationship between nanostructure and mechanical properties of  multicomponent systems and about the way in which these properties may be controlled. 

We would like to thank Dino Osmanovi{\'{c}}, Itay Azizi and David Kessler for helpful discussions. This work was supported by grants from the Israel Science Foundation and from the Israeli Centers for Research Excellance program of the Planning and Budgeting Committee.

\bibliography{references}

\end{document}